\documentclass[11pt]{article}
\usepackage{graphicx,wrapfig,amssymb,amsmath,multicol}
\usepackage{natbib}
\usepackage{hyperref}
\bibpunct{[}{]}{,}{n}{,}{,}
\usepackage{bm}
\usepackage{amsmath}
\graphicspath{{./figs/}{./png/}}

\title{\textbf{Turbulence and order in magnetized flowing plasmas}\\{\small{White paper for APS-DPP-CPP}}}
\author{Fatima Ebrahimi (PPPL/Princeton University) \\ \\ co-authors: Eric Blackman (Univ. of Rochester), Wendell Horton (Univ. of Texas)}
\date{December 2019}

\begin{document}

\maketitle
 Continued observational discoveries of high energy emission and large scale jets from compact accretion flows, including those from black hole engines, highlight the fundamental  role of \textbf{magnetized plasma} in  explaining many of the most luminous sources in the universe.  The engines of these high energy astrophysical accretors, along with  stars  and galaxies, also commonly show evidence for  the  contemporaneous
presence of both turbulence and magnetic fields ordered on spatial or temporal scales larger than those of
the fluctuations. Explaining this dichotomy  has been
a long standing challenge in basic plasma astrophysics across a range of sources.
\citep{Brandenburg2005,Blackman2015H} How such fields grow, given the presence  of fluctuations, what are the relative dynamical influences of large vs. small scale fields, what are the best analysis methods, and   
what minimum ingredients for growth are needed 
\citep{vishniac97,brandenburg2008,Yousef2008,Heinemann2011,rincon2011,Squire2015b, EB_2016, ebrahimi2019}, and 
the role of 
magnetic helicity transport
\cite{blackmanfield2002,vishniac_cho,bh_yuan95,ebrahimiprl} are all topics of active investigation.\\
\begin{wrapfigure}{r}{2.4in}
\vspace{-5mm}
\centering
\includegraphics[width=0.45\textwidth, height=0.25\textheight]{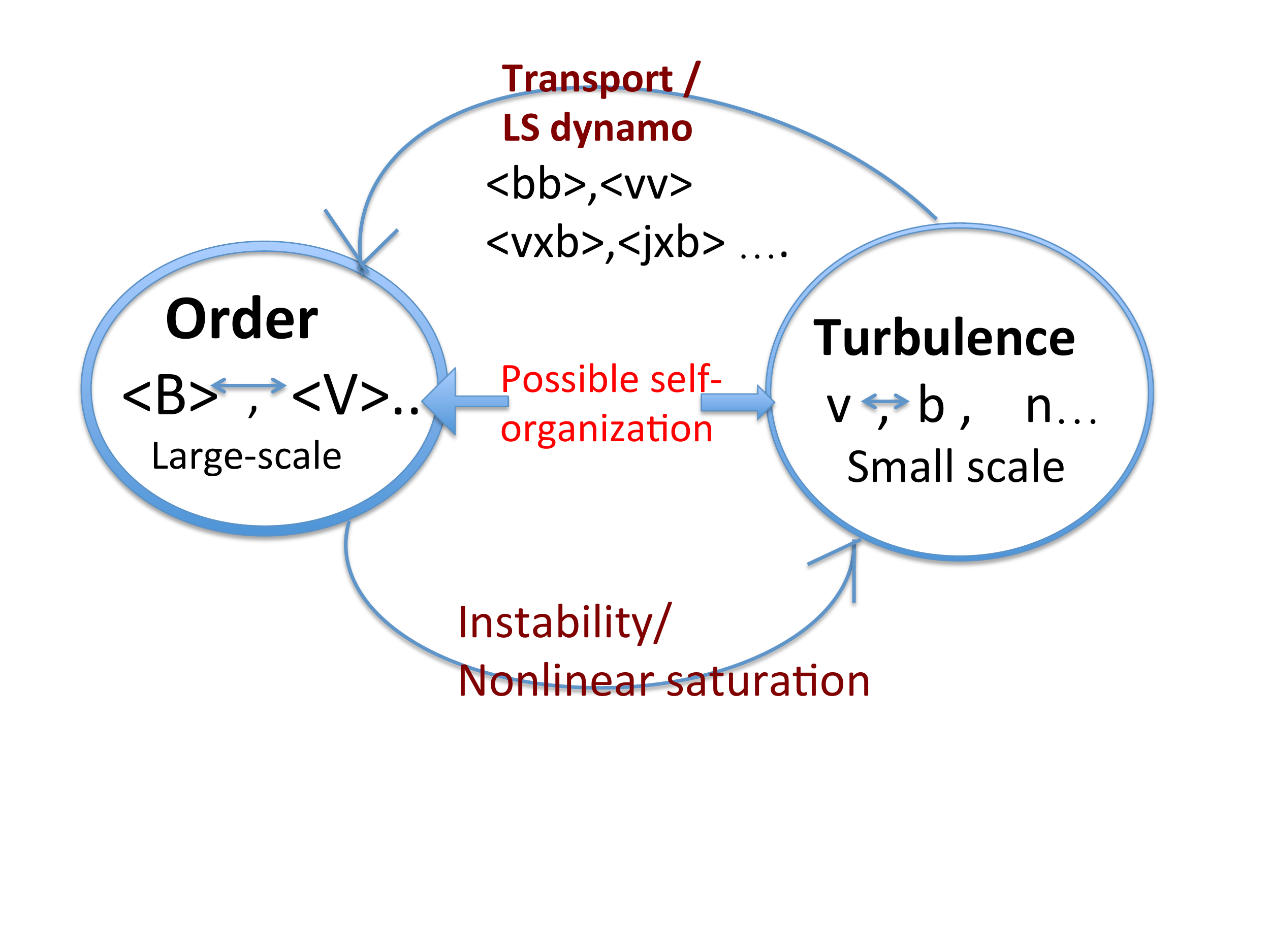}
\vspace{-22mm}
\caption{Several types of physical dynamics, including instabilities and nonlinear coupling, could link the large scales to small scales. Fields themselves are all coupled at all scales, as plasma motions and flows make magnetic fields, and magnetic field annihilation through reconnection could make MHD outflows (or accelerate particles).}
\label{fig:fields}
\end{wrapfigure}
Large-scale ordered fields seem to be coupled with  large-scale momentum transfer (in the form of jets and coronae)  in the otherwise likely turbulent  plasmas of astrophysical accretion disks.   Small and large scale  magnetic fields in astrophysical disks facilitate both the dynamics of angular momentum in disks via magnetorotational instability, as well as the formation and  collimation of jets, and creation of flux ropes and particle energization by reconnection  in disk coronae. The underlying physical mechanisms
for generation of large-scale fields in disks, and the radial and
vertical momentum transport, are interlinked. The magnetic stress spectrum produced  in the disk can  directly influence the fraction of energy dissipated within the disk (typically thermal) vs. above the disk (typically non-thermal).

Starting with any
seed field, the generation and amplification of fields  above some suitably defined forcing scale typical constitutes  a "large-scale dynamo (LSD)."   The turbulent amplification of fields  at or below the forcing scales is referred to as "small scale" or "fluctuation" dynamo (SSD) and can be contemporaneous with the LSD.\\

\textbf{\large{Turbulent small-scale vs large-scale dynamos:} }Small-scale (turbulent) dynamo is usually driven due to random stretching
of the magnetic field by the turbulent motions. This results in the generation of folded,
intermittent (but not volume filling) structures (see Fig.~\ref{fig:fields}).~\cite{schekochihin2004,haugen2004}.
The SSD is likely to be important 
for the early generation of magnetic fields in stars and galaxies/inter-stellar medium (ISM).~\cite{kulsrud1999}
Such SSD generated fields would then be present in
the plasma from stars or the ISM that could source accretion disks. LSDs however  require special conditions in the underlying flow turbulence
to be operational. The turbulence needs to  break 
mirror symmetry   for large scale dynamo action which can be accomplished by randomly forced helical turbulence  \cite{moffatt,Parker1979}, or large-scale shear flow with non-axisymmetric perturbation is needed~\cite{ebrahimi2019}.
\vspace{2mm}
\begin{wrapfigure}{r}{2.4in}
\vspace{+5mm}
\centering
\includegraphics[width=0.35\textwidth, height=0.2\textheight]{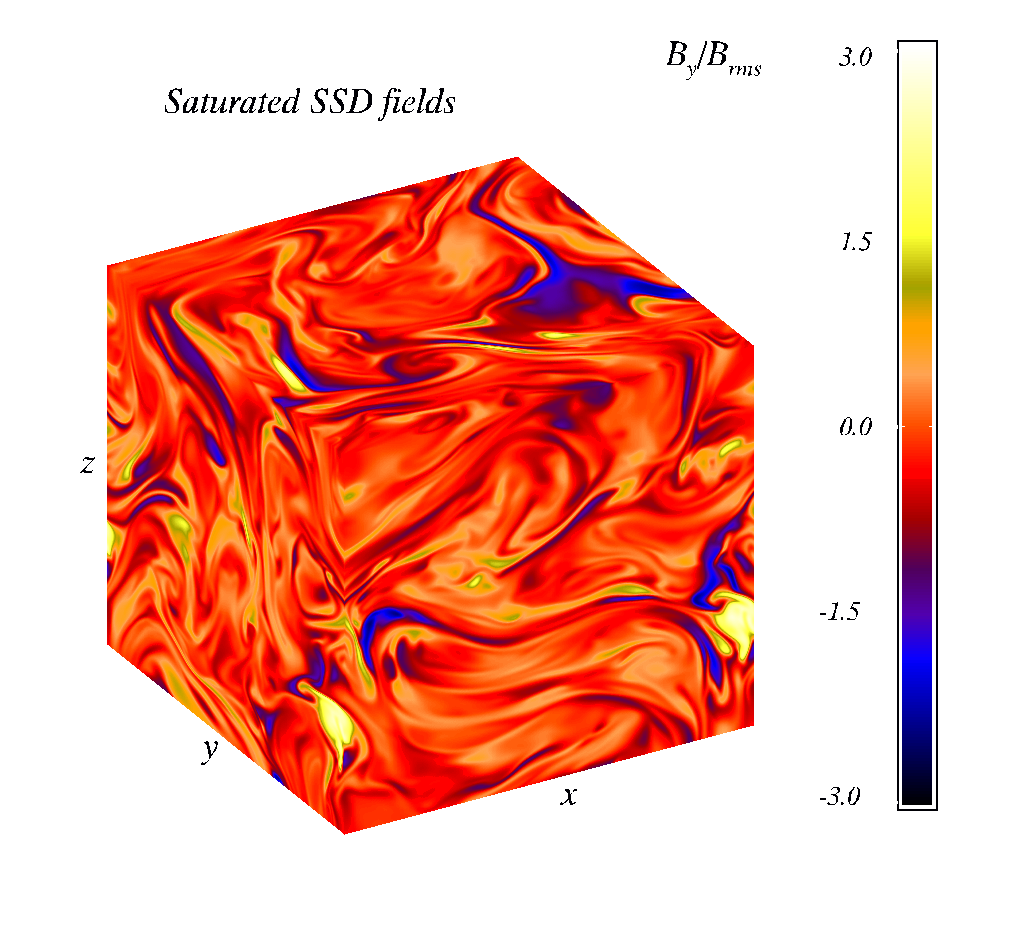}
\includegraphics[width=0.35\textwidth, height=0.2\textheight]{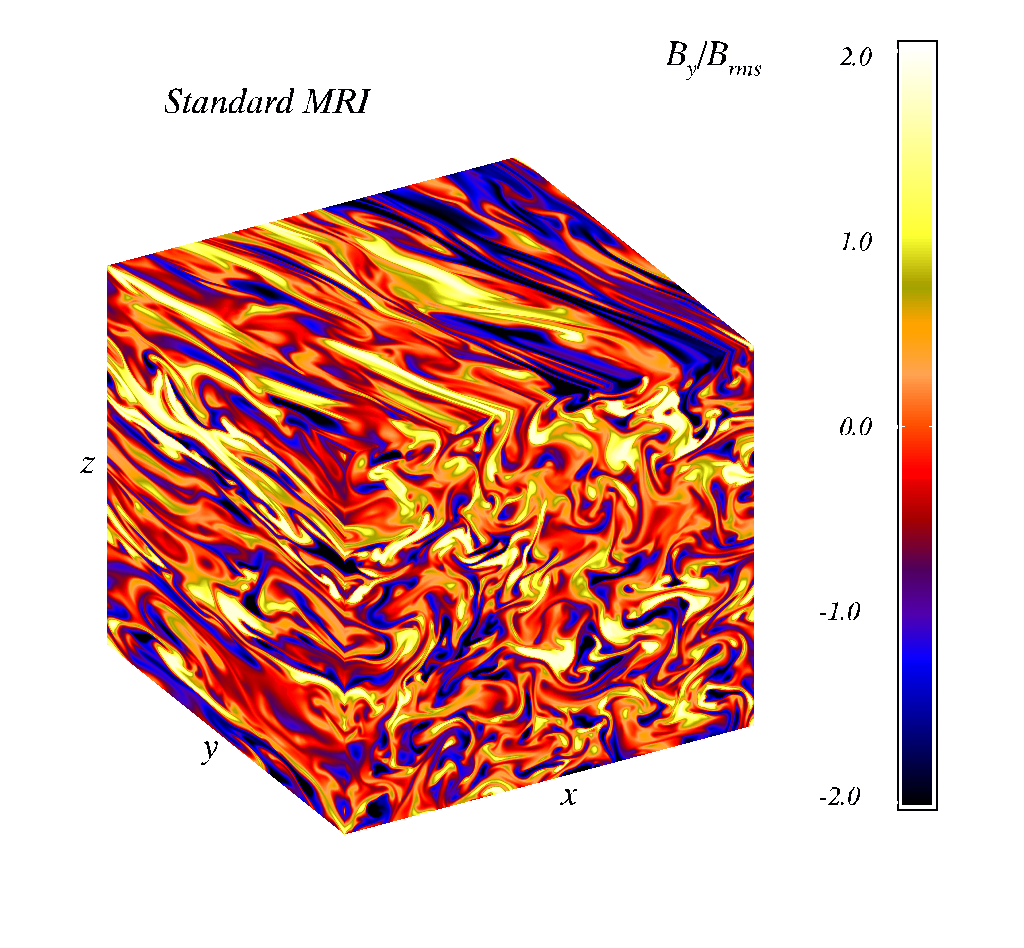}\\
\vspace{-7mm}
\caption{The azimuthal component of the magnetic fields obtained from numerical shearing box simulations  with explicit resistivity. Top: SSD with random non-helical forcing. Bottom: LSD fields from MRI simulations. Mean large-scale fields survive after planar averaging.~\cite{bhat2017,bhat2016}
}
\label{fig:fields}
\end{wrapfigure}
In  magnetically dominated flows of the laboratory, large-scale magnetic relaxation dynamos are  directly triggered by a 3-D instability whereby small scale helical fields relax to large scale helical fields. A similar instability-driven onset of LSD  also can occur in flow driven systems,  and the MRI dynamo is such an example.
Ordered fields resulting from LSDs are  known to cause~\cite{ebrahimi2009} or contribute to the saturation of flow-driven turbulent transport in MRI unstable shearing sheet flows~\citep{lesurdynamo,stratified2010,simon2011,Guan2011,Nauman2014}.     The direct dynamo action due the MRI non-axisymmetric mode itself  was first shown through quasilinear and computational calculations,~\cite{ebrahimi2009} and was demonstrated in 3-D global cylindrical simulations,~\cite{EB_2016} as well as the exponential growth of LS fields in the early phase of local shearing box simulations with explicit resistivity~\cite{bhat2016} was also revealed. This direct MRI LSD action has more recently been confirmed in simulations with a different code.~\cite{oishi2019}\\

\textbf{\large{Turbulent vs. ordered zonal flows:}}
Similar to the turbulent and ordered magnetic field structures, in flow-dominated plasmas, the flow itself  could exhibit and develop structures and correlated transport properties (such as  helicity) over a range of scales. In particular, nonuniform flows are  ubiquitous both in nature and in the laboratory: they occur in atmospheres, oceans, stars, galaxies, pipe flows, and magnetically confined plasmas, etc. The problem of the onset and self-sustenance of turbulence in spectrally stable nonuniform/shear flows is a challenge for fluid/plasma dynamics research. In such flows, perturbations (of certain spatial characteristics) undergo only linear transient growth leading to short perturbation life times. The imperfect linear growth must be compensated by  nonlinear positive feedback to repopulate  transiently growing perturbations. Subtle interplay of the linear transient and nonlinear processes can self-organize (chaotic or coherent) perturbations and ensure their self-sustenance. In short, the dynamical interaction of all scales, through the properties of turbulent cascades~\cite{horton2010,gogichaishvili2017,gogichaishvili2018}, as well as the generated ordered flows ~\cite{mynavier,parker14,parker2019,fraser2017} are essential for understanding the structures and properties
%
of  turbulent flows.

A variety of computational techniques and a hierarchy of physics models are needed to understand  turbulence and the growth of ordered fields in  flow-dominated astrophysical plasmas. Although the validity of physical models (collisional vs. collisionless) depends on the astrophysical context, there
remains a need to compare
basic  physics models 
with converged numerical simulations in order to better inform practical models of e.g. accretion disks that can be used to compare with observations.

Due to the multi-scale nature of most flow-dominated astrophysical settings, a hierarchy of approaches is  necessary and often beneficial.  However,  it is also possible to get  locked into the limitations of a given tool which can bias or limit our understanding. For example, focusing on 
specifically "local" disk models
has one set of limitations whilst
fully "global" simulations has another. The same applies to  specifically  "multi-fluid" plasma models vs. say "gyro-kinetic" models.

For discovery through theory and computations, code-code comparison, physics model comparisons (for example varying  ion Larmor radius in kinetic models and two-fluid models to find an overlap for these two models)  should therefore be strongly encouraged. Below we discuss several examples where 
diverse computational approaches
can be fruitful.
\\
\\

\textbf{\large{1- Global vs. local computational models:}}
Much of the existing theoretical and numerical studies of the saturation mechanism of flow-driven
turbulent media (and dynamo) are based on either (1) simple local approximations, such as shearing-box
simulations, or (2) global simulations aimed toward real astrophysical systems. But intermediate  complementary numerical approaches (without the complexity of full global simulations or the limitation of the
shearing-box model) can be essential. The effect of boundaries (periodic vs. free or impenetrating boundaries) and simulation sizes (box size, for example), can be investigated. Such studies would help to identify the underlying physics and  any possible dependence of physics results on the numerical  method/domain.

\textbf{\large{2- The need for hierarchy of physics models:}}
The enormous scale separation in astrophysics, as well as wide range of physical parameters (for example magnetic Reynolds number) necessitates a hierarchy of physical treatments for astrophysical settings.  Basic models ranging from collisional to collisionless models, including single and two-fluid, particle, hybrid,  and continuum
Vlasov-Maxwell (and with including atomic and molecular,
radiation, general relativity) should all be investigated.
To assess the validity of the results, physics models should be validated against observational and experimental data. The correctness of the complex models should also be verified with reduced analytical models.  Basic theory, code-code comparison, and validation against existing observational and experimental data, are the key to uncover the underlying physics in complex and multi-scale flow-dominated systems. \\


{\small
\setlength{\baselineskip}{0.9\baselineskip}
\vspace{-.25in}
\bibliographystyle{unsrtnat} 

\setlength{\baselineskip}{1.1\baselineskip}
}
\end{document}